\documentclass[
    letter,
    center,
    oneside,
    final,
    nocrop,
    notheoremsty
]{NAR}

\usepackage[T1]{fontenc}
\usepackage{helvet}
\usepackage[numbers,round,sort&compress]{natbib}
\renewcommand{\bibfont}{\fontsize{8}{9}\selectfont}
\makeatletter
\def\@biblabel#1{#1.}
\makeatother
\newcommand{\skipcite}[1]{}

\PassOptionsToPackage{hyphens}{url}
\usepackage{xcolor}
\definecolor{citecolor}{rgb}{0,0,1}
\definecolor{urlcolor}{rgb}{0,0,1}
\definecolor{linkcolor}{rgb}{0,0,1}
\usepackage[hyperfootnotes, breaklinks=true, colorlinks=true,
urlcolor=urlcolor, linkcolor=linkcolor, citecolor=citecolor, bookmarks=true, bookmarksdepth=2]{hyperref}%

\makeatletter
\@namedef{toclevel@article}{0}
\makeatother

\usepackage{scalerel}
\usepackage{tikz}
\usetikzlibrary{svg.path}

\definecolor{orcidlogocol}{HTML}{A6CE39}
\tikzset{
  orcidlogo/.pic={
    \fill[orcidlogocol] svg{M256,128c0,70.7-57.3,128-128,128C57.3,256,0,198.7,0,128C0,57.3,57.3,0,128,0C198.7,0,256,57.3,256,128z};
    \fill[white] svg{M86.3,186.2H70.9V79.1h15.4v48.4V186.2z}
                 svg{M108.9,79.1h41.6c39.6,0,57,28.3,57,53.6c0,27.5-21.5,53.6-56.8,53.6h-41.8V79.1z M124.3,172.4h24.5c34.9,0,42.9-26.5,42.9-39.7c0-21.5-13.7-39.7-43.7-39.7h-23.7V172.4z}
                 svg{M88.7,56.8c0,5.5-4.5,10.1-10.1,10.1c-5.6,0-10.1-4.6-10.1-10.1c0-5.6,4.5-10.1,10.1-10.1C84.2,46.7,88.7,51.3,88.7,56.8z};
  }
}

\newcommand\orcid[1]{\href{https://orcid.org/#1}{\raisebox{4pt}[0pt][0pt]{\scalerel*{
\begin{tikzpicture}[yscale=-1,transform shape]
\path (0,0) pic{orcidlogo};
\end{tikzpicture}
}{+}}}}

\usepackage{lineno}
\copyrightyear{2022}
\pubdate{XX XXX 2022}
\pubyear{2022}
\jvolume{50}
\jissue{Web server issue}

\makeatletter
\def\ps@plain{\let\@mkboth\@gobbletwo
     \def\@oddhead{%
     }%
     \def\@evenhead{}%
      \def\@oddfoot{}
      \def\@evenfoot{}
     }
\def\ps@myheadings{%
      \let\@oddfoot\@empty
      \let\@evenfoot\@empty
      \def\@evenhead{}%
      \def\@oddhead{}%
    \let\@mkboth\@gobbletwo
    \let\chaptermark\@gobble
    \let\sectionmark\@gobble
    \def\titlemark##1{\markboth{##1}{##1}}%
    \def\authormark##1{\gdef\leftmark{##1}}%
}
\def\ps@plain{\let\@mkboth\@gobbletwo
     \def\@oddhead{}%
     \def\@evenhead{}%
      \def\@oddfoot{}
      \def\@evenfoot{}
     }

\renewcommand\section{\@startsection {section}{1}{\z@}%
                                   {-8pt \@plus -2\p@ \@minus -2\p@}%
                                   {5pt}%
                                   {\sectionfont}}
\renewcommand\subsection{\@startsection{subsection}{2}{\z@}%
                                   {-4pt \@plus -2\p@ \@minus -2\p@}%
                                   {5\p@}%
                                   {\subsectionfont}}
\def\catchlineinfo{\ifx\@jname\empty\else\begin{tabular}[t]{@{}r@{}}
\@jname,\ \@pubyear,\ \ifx\@jvol\undefined\else Vol.\ \@jvol,\ \fi\ifx\@jissue\undefined\else \@jissue\hskip1em\fi\ \textbf{\textit{D\@firstpage--D\@lastpage}}
\end{tabular}\fi}
\makeatother

\renewcommand{\nocite}[1]{}

\begin{document}
\thispagestyle{plain}
\title{
BioSimulators: a central registry of simulation engines and services for recommending specific tools}

\author{%
Bilal Shaikh\orcid{0000-0001-5801-5510}$^\text{\sfb 1}$,
Lucian P. Smith\orcid{0000-0001-7002-6386}$^\text{\sfb 2}$,
Dan Vasilescu\orcid{}$^\text{\sfb 3}$,
Gnaneswara Marupilla\orcid{}$^\text{\sfb 3}$,
Michael Wilson\orcid{0000-0001-5892-6074}$^\text{\sfb 3}$,
Eran Agmon\orcid{0000-0003-1279-2474}$^\text{\sfb 4}$,
Henry Agnew\orcid{0000-0003-1447-6045}$^\text{\sfb 5}$,
Steven S. Andrews\orcid{0000-0002-4576-8107}$^\text{\sfb 2}$,
Azraf Anwar\orcid{}$^\text{\sfb 6}$,
Moritz E. Beber\orcid{0000-0003-2406-1978}$^\text{\sfb 7}$,
Frank T. Bergmann\orcid{0000-0001-5553-4702}$^\text{\sfb 8}$,
David Brooks\orcid{0000-0002-6758-2186}$^\text{\sfb 9}$,
Lutz Brusch\orcid{0000-0003-0137-5106}$^\text{\sfb 10}$,
Laurence Calzone\orcid{0000-0002-7835-1148}$^\text{\sfb 11}$,
Kiri Choi\orcid{0000-0002-0156-8410}$^\text{\sfb 12}$,
Joshua Cooper\orcid{}$^\text{\sfb 13}$,
John Detloff\orcid{}$^\text{\sfb 14}$,
Brian Drawert\orcid{0000-0002-0543-8189}$^\text{\sfb 13}$,
Michel Dumontier\orcid{0000-0003-4727-9435}$^\text{\sfb 15}$,
G. Bard Ermentrout\orcid{0000-0002-5854-0654}$^\text{\sfb 16}$,
James R. Faeder\orcid{0000-0001-8127-609X}$^\text{\sfb 16}$,
Andrew P. Freiburger\orcid{0000-0002-7288-535X}$^\text{\sfb 17}$,
Fabian Fröhlich\orcid{0000-0002-5360-4292}$^\text{\sfb 18}$,
Akira Funahashi\orcid{0000-0003-0605-239X}$^\text{\sfb 19}$,
Alan Garny\orcid{0000-0001-7606-5888}$^\text{\sfb 9}$,
John H. Gennari\orcid{0000-0001-8254-4957}$^\text{\sfb 20}$,
Padraig Gleeson\orcid{0000-0001-5963-8576}$^\text{\sfb 21}$,
Anne Goelzer\orcid{0000-0003-2222-6142}$^\text{\sfb 22}$,
Zachary Haiman\orcid{0000-0001-6175-5050}$^\text{\sfb 23}$,
Joseph L. Hellerstein\orcid{0000-0003-0802-4069}$^\text{\sfb 2}$,
Stefan Hoops\orcid{0000-0001-8503-8371}$^\text{\sfb 24}$,
Jon C. Ison\orcid{0000-0001-6666-1520}$^\text{\sfb 25}$,
Diego Jahn\orcid{0000-0001-6774-5507}$^\text{\sfb 10}$,
Henry V. Jakubowski\orcid{0000-0002-9629-9339}$^\text{\sfb 26}$,
Ryann Jordan\orcid{}$^\text{\sfb 1}$,
Matúš Kalaš\orcid{0000-0002-1509-4981}$^\text{\sfb 27}$,
Matthias König\orcid{0000-0003-1725-179X}$^\text{\sfb 28}$,
Wolfram Liebermeister\orcid{0000-0002-2568-2381}$^\text{\sfb 22}$,
Synchon Mandal\orcid{0000-0002-1212-5279}$^\text{\sfb 29}$,
Robert McDougal\orcid{0000-0001-6394-3127}$^\text{\sfb 30}$,
J. Kyle Medley\orcid{0000-0002-1509-4981}$^\text{\sfb 31}$,
Pedro Mendes\orcid{0000-0001-6507-9168}$^\text{\sfb 3}$,
Robert Müller\orcid{}$^\text{\sfb 10}$,
Chris J. Myers\orcid{0000-0002-8762-8444}$^\text{\sfb 32}$,
Aurelien Naldi\orcid{0000-0002-6495-2655}$^\text{\sfb 33}$,
Tung V. N. Nguyen\orcid{0000-0002-2876-6046}$^\text{\sfb 34}$,
David P. Nickerson\orcid{0000-0003-4667-9779}$^\text{\sfb 9}$,
Brett G. Olivier\orcid{0000-0002-5293-5321}$^\text{\sfb 35}$,
Drashti Patoliya\orcid{}$^\text{\sfb 36}$,
Loïc Paulevé\orcid{0000-0002-7219-2027}$^\text{\sfb 37}$,
Linda R. Petzold\orcid{0000-0001-6251-6078}$^\text{\sfb 38}$,
Ankita Priya\orcid{}$^\text{\sfb 39}$,
Anand K. Rampadarath\orcid{0000-0001-8830-6212}$^\text{\sfb 9}$,
Johann M. Rohwer\orcid{0000-0001-6288-8904}$^\text{\sfb 40}$,
Ali S. Saglam\orcid{0000-0002-6513-8401}$^\text{\sfb 16}$,
Dilawar Singh\orcid{0000-0002-4645-3211}$^\text{\sfb 41}$,
Ankur Sinha\orcid{0000-0001-7568-7167}$^\text{\sfb 42}$,
Jacky Snoep\orcid{0000-0002-0405-8854}$^\text{\sfb 40}$,
Hugh Sorby\orcid{0000-0001-8991-4703}$^\text{\sfb 9}$,
Ryan Spangler\orcid{}$^\text{\sfb 43}$,
Jörn Starruß\orcid{0000-0003-3649-2433}$^\text{\sfb 10}$,
Payton J. Thomas\orcid{0000-0002-5075-3911}$^\text{\sfb 44}$,
David van Niekerk\orcid{0000-0003-2200-7935}$^\text{\sfb 40}$,
Daniel Weindl\orcid{0000-0001-9963-6057}$^\text{\sfb 45}$,
Fengkai Zhang\orcid{0000-0001-7112-9328}$^\text{\sfb 46}$,
Anna Zhukova\orcid{0000-0003-2200-7935}$^\text{\sfb 47}$,
Arthur P. Goldberg\orcid{0000-0003-2772-1484}$^\text{\sfb 1}$,
Michael L. Blinov\orcid{0000-0002-9363-9705}$^\text{\sfb 3}$,
Herbert M. Sauro\orcid{0000-0002-3659-6817}$^\text{\sfb 2}$,
Ion I. Moraru\orcid{0000-0002-3746-9676}$^\text{\sfb 3}$
Jonathan R. Karr\orcid{0000-0002-2605-5080}$^\text{\sfb 1,*}$%
\footnote{To whom correspondence should be addressed.
Tel: +1 212-659-8973 (JRK); Email: \href{mailto:karr@mssm.edu}{karr@mssm.edu}}}

\address{%
$^\textsf{1}$Icahn School of Medicine at Mount Sinai, New York, NY 10029, US,
$^\textsf{2}$University of Washington, Seattle, WA 98105, US,
$^\textsf{3}$University of Connecticut School of Medicine, Farmington, CT 06030, US,
$^\textsf{4}$Stanford University, Stanford, CA 94305, US,
$^\textsf{5}$LibreTexts, US,
$^\textsf{6}$New York University, Brooklyn, NY 11201,
$^\textsf{7}$Unseen Bio ApS, 2100 København Ø, DK,
$^\textsf{8}$Heidelberg University, 69120 Heidelberg, DE,
$^\textsf{9}$University of Auckland, 1010 Auckland, NZ,
$^\textsf{10}$Technical University of Dresden, 01187 Dresden, DE,
$^\textsf{11}$Institut Curie, 75248 Paris, FR,
$^\textsf{12}$Korea Institute for Advanced Study, 02455 Seoul, KR,
$^\textsf{13}$University of North Carolina, Asheville, Ashville, NC 28804, US,
$^\textsf{14}$Independent, Madison, WI 53705, US,
$^\textsf{15}$Maastricht University, 6200 Maastricht, NL,
$^\textsf{16}$University of Pittsburgh, Pittsburgh, PA 15260, US,
$^\textsf{17}$University of Victoria, Victoria, BC V8P 5C2, CA,
$^\textsf{18}$Harvard Medical School, Boston, MA 02115 US,
$^\textsf{19}$Keio University, Yokohama 223-8522, JP,
$^\textsf{20}$University of Washington, Seattle WA 98019, US,
$^\textsf{21}$University College London, London WC1E 6BT, UK,
"$^\textsf{22}$National Research Institute for Agriculture, Food and Environment\\
Université Paris-Saclay, 78350 Jouy-en-Josas, FR,"
$^\textsf{23}$University of California, San Diego, La Jolla, CA 92093, US,
$^\textsf{24}$University of Virginia, Charlottesville, VA 22904, US,
$^\textsf{25}$CNRS, UMS 3601, Institut Français de Bioinformatique, IFB-core, 91000 Évry-Courcouronnes, FR,
$^\textsf{26}$College of Saint Benedict and Saint John’s University, St. Joseph, MN 56374, US,
$^\textsf{27}$University of Bergen, 5020 Bergen, NO,
$^\textsf{28}$Humboldt University of Berlin, 10115 Berlin, DE,
$^\textsf{29}$Technical University of Dresden, 01069 Dresden, DE,
$^\textsf{30}$Yale University, New Haven, CT 06511, US,
$^\textsf{31}$Autodesk, Inc., San Rafael, CA 94903, US,
$^\textsf{32}$University of Colorado at Boulder, Boulder CO, 80309, US,
$^\textsf{33}$Inria Saclay - Île-de-France Research Centre, 91120 Palaiseau, FR,
$^\textsf{34}$European Molecular Biology Laboratory - European Bioinformatics Institute, Hinxton, Cambridge CB10 1SD, UK,
$^\textsf{35}$Vrije Universiteit Amsterdam, 1081 HZ Amsterdam, NL,
$^\textsf{36}$Sarvajanik College of Engineering \& Technology, Surat, Gujarat 395001, IN,
$^\textsf{37}$Centre National de la Recherche Scientifique, 33400 Talence, France,
$^\textsf{38}$University of California, Santa Barbara, Santa Barbara, CA 93106, US,
$^\textsf{39}$Birla Institute of Technology, Mesra, Jharkhand 835215, IN,
$^\textsf{40}$Stellenbosch University, Stellenbosch, 7600, ZA,
$^\textsf{41}$Subconscious Compute Pvt. Ltd., Bangalore, IN,
$^\textsf{42}$University College London, London, WC1E 6BT, UK,
$^\textsf{43}$Allen Institute for Cell Science, Seattle, WA 98109, US,
$^\textsf{44}$University of Utah, Salt Lake City, UT 84112, US,
$^\textsf{45}$Helmholtz Zentrum München GmbH and German Research Center for Environmental Health, 85764 Neuherberg, DE,
$^\textsf{46}$National Institutes of Health, Bethesda, MD 20892, US and
$^\textsf{47}$Institut Pasteur, 75015 Paris, FR.
}


\maketitle

\clearpage

\begin{abstract}
Computational models have great potential to accelerate bioscience, bioengineering, and medicine. However, it remains challenging to reproduce and reuse simulations, in part, because the numerous formats and methods for simulating various subsystems and scales remain siloed by different software tools. For example, each tool must be executed through a distinct interface. To help investigators find and use simulation tools, we developed BioSimulators (\href{https://biosimulators.org}{https://\allowbreak{}bio\allowbreak{}sim\allowbreak{}u\allowbreak{}la\allowbreak{}tors.\allowbreak{}org}), a central registry of the capabilities of simulation tools and consistent Python, command-line, and containerized interfaces to each version of each tool. The foundation of BioSimulators is standards, such as CellML, SBML, SED-ML, and the COMBINE archive format, and validation tools for simulation projects and simulation tools that ensure these standards are used consistently. To help modelers find tools for particular projects, we have also used the registry to develop recommendation services. We anticipate that BioSimulators will help modelers exchange, reproduce, and combine simulations.
\end{abstract}

\enlargethispage{0pt}%

\section{Introduction}
Sophisticated computational models that can predict biological phenomena have great potential for bioscience, bioengineering, and medicine. For example, whole-cell models could help scientists understand the origin of behavior, help engineers design biofactories, and help clinicians personalize medicine \cite{carrera2015build, marucci2020computer}. Due to the complexity of biology, such models often need to integrate multiple subsystems across multiple scales, requiring collaborations among teams and the use of multiple tools \cite{szigeti2018blueprint, waltemath2016toward}. 

Over the last 25 years, researchers have developed numerous methods and tools for simulating various subsystems and scales. For example, COBRApy \cite{ebrahim2013cobrapy} and COPASI \cite{bergmann2017copasi} can execute constraint-based and kinetic simulations of metabolic and signaling networks, respectively.

Toward combining simulations, the community has developed several resources for sharing several types of models. For example, formats such as CellML \cite{clerx2020cellml} and SBML \cite{keating2020sbml}; libraries for these formats such as lib\-Cell\-ML and libSBML\nocite{bornstein2008libsbml}; and repositories such as BioModels \cite{malik2020biomodels} and ModelDB \cite{mcdougal2017twenty} help investigators share and reuse diverse types of models.

More recently, investigators have initiated similar efforts to share several types of simulations. For example, the Simulation Experiment Description Language (SED-ML; \citenum{Smith2021}), the
COMBINE archive format \cite{bergmann2014combine}, the Kinetic Simulation Algorithm Ontology (KiSAO; \citenum{courtot2011controlled}), and the JWS Online repository \cite{peters2017jws} can be used to share kinetic simulations.

Despite this progress, sharing, reusing, and combining simulations remains difficult. One reason is that it is difficult to find, obtain, and use appropriate tools for particular systems and scales. For example, many tools do not provide clear documentation about their simulation methods, and each tool must be obtained from a different location, installed via a different process, and executed via a different UI or API using different model formats. Guides, such as the retired SBML Software Guide, and container registries, such as BioContainers\nocite{da2017biocontainers}, have only addressed some of these issues.

To accelerate the reuse of simulations, as well as the development of multiscale simulations, we developed BioSimulators, a central registry for the capabilities of simulation tools (e.g., supported model formats, modeling frameworks, and simulation algorithms) and consistent Python, command-line, and containerized interfaces to these tools. Currently, BioSimulators encompasses 54 tools for 13 model formats, 14 modeling frameworks, and 91 simulation algorithms (\textcolor{citecolor}{Tables~S1-S3}), and consistent interfaces to 21 of these tools (\textcolor{citecolor}{Tables~S4,S5}). For example, this includes asynchronous logical simulation with BoolNet, geometric flux balance analysis with COBRApy, discrete particle-based simulation with BioNetGen, and discrete spatial simulation with Smoldyn. To help investigators find appropriate tools, we have also used this registry to develop services for recommending specific algorithms and tools for particular systems and scales.

\begin{figure}[t!]
    \centering
    \includegraphics[width=\columnwidth]{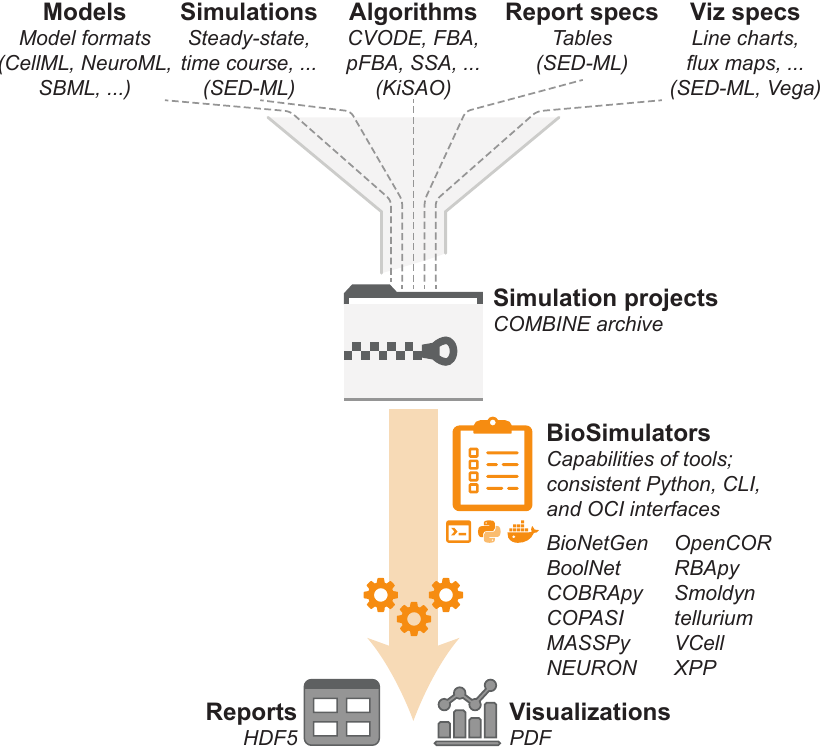}
    \caption{\textbf{BioSimulators simplifies simulation by abstracting simulation projects and simulation tools.} BioSimulators abstracts projects as COMBINE archives and tools as containerized command-line interfaces. These abstractions make it easier to execute a broad range of simulations.
    \vspace{-20pt}}
    \label{fig:overview}
\end{figure}

\begin{figure*}[t!]
    \centering
    \includegraphics[width=\textwidth]{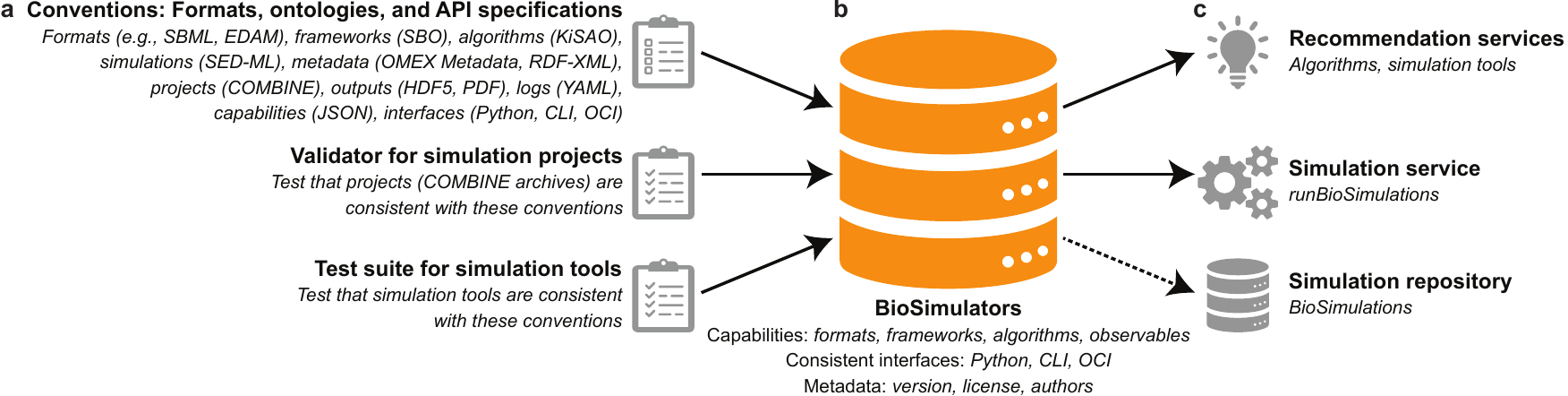}
    \caption{\textbf{Overview of the BioSimulators ecosystem.} The foundation of BioSimulators (\textbf{b}) is an integrated set of formats, ontologies, and specifications for simulation projects and simulation tools, and tools for checking that these conventions are used consistently (\textbf{a}). These conventions make it easier to work with multiple types of simulations. To further help investigators find and run simulation tools, we have also developed user-friendly services for recommending tools, executing simulations, and visualizing the results of simulations. In addition, we are developing a repository for sharing projects, their results, and visualizations of these results (\textbf{c}).
    \vspace{-10pt}
    }
    \label{fig:ecosytem}
\end{figure*}

To simplify the discovery, installation, and use of simulation tools, BioSimulators is based on an integrated set of formats, ontologies, and quality controls (\autoref{fig:overview}). BioSimulators uses Docker images to encapsulate simulation tools and a new schema to capture their capabilities. The input to each tool is a COMBINE archive which contains SED-ML files that describe simulations of models in formats such as SBML with algorithms described using KiSAO. The outputs of each tool are HDF5\nocite{folk2011overview} and PDF files that capture data sets and visualizations of simulation results. To ensure these resources are used consistently, we also developed tools for integrated validation of simulation projects and tools (\autoref{fig:ecosytem}\textcolor{citecolor}{a}).
On top of BioSimulators, we have also developed runBioSimulations and BioSimulations, user-friendly web applications for using BioSimulators to execute and share simulations and visualizations of their results \cite{shaikh2021runbiosimulations} (\autoref{fig:ecosytem}\textcolor{citecolor}{c}).

Below, we summarize the key features of BioSimulators, describe its architecture, delineate several use cases for BioSimulators, and outline the future directions of BioSimulators. Tutorials and additional documentation are available at \href{https://docs.biosimulations.org}{https://\allowbreak{}docs.\allowbreak{}bio\allowbreak{}sim\allowbreak{}u\allowbreak{}la\allowbreak{}tions.\allowbreak{}org}.

\section{Key features}
The central features of BioSimulators are streamlined abilities to find, obtain, and use simulation tools for a broad range of modeling frameworks, formats, and algorithms.

\subsection{Streamlined discovery of appropriate tools for projects}
To help investigators find appropriate tools, BioSimulators provides a central database of the capabilities of simulation tools. This includes the model formats, modeling frameworks, simulation algorithms, and observables that each tool supports; the parameters of each algorithm; and the data type of each parameter, as well as metadata, such as the license of each tool. Where possible, this information is captured using ontologies such as EDAM, KiSAO, SBO, and SIO. This ensures that simulation tools are described consistently. 

To further help investigators find tools, we have also used the capabilities of each tool and relationships among algorithms captured by KiSAO to develop recommendation services. For example, we have developed a web form that can recommend simulators for executing particular projects.

\subsection{Streamlined acquisition and installation of simulators}
To make it easier to obtain and install simulation tools, BioSimulators saves a Docker image for each version of each containerized tool. This ensures that investigators can use a single program, such as Docker Desktop, to easily obtain and install any version of any tool. To ensure that investigators can use these images in high-performance computing (HPC) environments, which generally disallow the use of Docker due to security limitations, BioSimulators tests that these images are compatible with the Singularity Image Format (SIF), which can be run in HPC environments. Similarly, the Python APIs and command-line programs for simulation tools can be installed consistently from the PyPI repository.

\subsection{Streamlined execution of simulation tools}
To simplify simulation, each containerized tool provides the same command-line interface. These interfaces capture the project to be executed and the location where its outputs (reports and visualizations) should be saved. This enables modelers to use multiple simulators simply by switching their image ids. We anticipate this will help investigators work with a broader range of simulations, especially trainees, experimentalists, and peer reviewers. Most of the simulators with containerized interfaces also provide Python APIs. These APIs provide consistent, flexible low-level simulation capabilities. Currently, we are helping several groups use these APIs to develop interactive tools for research and education.

\subsection{Accurate and up-to-date information about simulators}
To ensure that the interfaces to simulators are consistent and that their specifications are accurate, we extensively review each version of each tool. The first version of each tool submitted to BioSimulators is both automatically validated by a test suite that we developed and manually reviewed by our team. Each subsequent version of each tool is also validated by this test suite. This test suite uses the simulator to execute a set of example COMBINE archives and checks that the tool produces the expected results. The test suite uses the specifications for the tool to select appropriate example archives. To enable developers to keep BioSimulators up to date, we provide an API that developers can use to automatically submit each version of their tools. We anticipate that this approach will enable us to keep BioSimulators up-to-date and accurate with minimal effort.

\section{Methods}
\subsection{Consistent representation of simulation projects and tools}
The foundation of BioSimulators is a set of formats, ontologies, and specifications for consistent representation of simulation projects (one or more simulations of one or more models using one or more algorithms), the inputs (e.g., experimental data for validating simulations) and outputs (data sets and visualizations of simulation results) of simulation projects, simulation tools, and the capabilities of simulation tools (supported formats, frameworks, and algorithms). Where possible, these conventions embrace existing resources. Using these resources required filling numerous gaps within and between them. This included creating new schemas for simulation results, logs of the execution of simulations, and the capabilities of simulation tools; formalizing numerous aspects of SED-ML; adding many new ontology concepts for formats and algorithms; and correcting hundreds of bugs in various simulation projects and software tools.

BioSimulators uses the COMBINE archive format to encapsulate all of the files that constitute a simulation project. Within COMBINE archives, BioSimulators uses formats such as BNGL, CellML, GINML, NeuroML/LEMS, RBA XML, SBML, Smoldyn, VCML, and the XPP ODE format to describe models and SED-ML to describe analyses of these models, such as simulations of time courses and steady-states. Within SED-ML, BioSimulators uses the KiSAO ontology to describe the algorithms and algorithm parameters for these analyses. To enable investigators to describe a broad range of simulations, we significantly expanded the KiSAO ontology and filled several gaps in SED-ML \cite{Smith2021}.

To consistently capture the outputs of simulation projects, we developed schemas for encoding the results of simulations into HDF5 files and encoding logs of the execution of simulation projects into YAML. We use the PDF format to capture visualizations of simulation results.

To enable modelers to execute simulators consistently, we developed specifications for Python APIs and containerized command-line programs for simulators. To help investigators find specific tools for particular projects, we developed a schema for capturing the capabilities of simulation tools. This schema uses the EDAM, SBO, KiSAO, SIO, and other ontologies to capture the model formats, modeling frameworks, simulation algorithms, and simulation observables that each tool supports. We similarly helped expand these ontologies to better capture the capabilities of a broader range of tools.

More information about these conventions is available in \textcolor{citecolor}{Section~S2} and at \url{https://docs.biosimulations.org}.

\subsection{Standardized interfaces to simulation tools}
We developed most of the Python, command-line, and containerized interfaces to simulation tools by wrapping simulation tools, such as COBRApy and COPASI, with BioSimulators-utils, a library that we developed for orchestrating the execution of COMBINE archives. Briefly, BioSimulators-utils executes each simulation task in each SED-ML file in a COMBINE archive by (1)~resolving the model for the task, (2)~modifying the model according to the changes specified in SED-ML, (3)~using KiSAO to determine the most similar simulation algorithm that the simulation tool implements to the algorithm specified for the task, (4)~translating this algorithm and its specified parameters into the corresponding method of the simulation tool and its arguments, (5)~executing this method with these arguments, (6)~collecting the results of this method, and (7)~using these results to generate the reports and plots specified in the SED-ML files. This modular design minimizes the effort needed to create standardized interfaces to simulation tools. The use of KiSAO to automatically identify suitable alternative algorithms enables investigators to both use SED-ML to precisely record the algorithms they used to execute simulations with one tool and execute the SED-ML files with additional tools that implement similar algorithms. \textcolor{citecolor}{Section~S4} contains more information about BioSimulators-utils.

\subsection{Recommendations of algorithms and simulation tools}
To help investigators navigate the sea of simulation formats, methods, and tools, we developed several interfaces for recommending resources, including (a)~an interactive table for searching our registry of tools; (b)~a web form for obtaining a list of tools which implement algorithms similar to a given algorithm, sorted by the maximal similarity of their algorithms to the given algorithm; and (c)~a web form for identifying simulators which can execute a given project using the specified or similar algorithms. Briefly, we implemented these services by (a)~determining the formats and algorithms specified for a given project, (b)~using our registry to determine the capabilities of each tool, (c)~using parent-child and other relationships to encode similarities among algorithms into KiSAO, (d)~using these relationships to query KiSAO for sets of similar algorithms, (e)~manually assigning each set a degree of similarity, (f)~combining the formats and algorithms required for a given project, the capabilities of each tool, and the similarity among algorithms to determine the maximal degree of similarity at which each tool can execute a given project, and (g)~sorting the tools by this maximal similarity. More information is available in \textcolor{citecolor}{Section~S4}.

\subsection{Validation of simulation projects and tools}
\label{projectValidation}
To ensure that BioSimulators' conventions are used consistently and to quickly alert users to issues, we developed a tool for integrated validation of COMBINE archives (model, SED-ML, and metadata files) and tools for validating simulation results, logs of the execution of simulations, and the capabilities of simulators described with the new schemas outlined above. This included developing the first validation rules for SED-ML. For example, our tool for validating simulation projects checks that each SED-ML file is consistent with the SED-ML schema and that each observable of each simulation references a valid model variable. To make these validation tools easy to use, we developed several interfaces, including web forms, a REST API, a command-line program, and a Python API. Four model repositories are already using these tools to debug their models and simulations.

Similarly, we also developed a test suite for checking whether simulation tools execute projects consistently with BioSimulators' conventions. Briefly, the test suite executes simulation tools with a set of test COMBINE archives and checks that they produced the expected outputs. These test archives enable the test suite to probe support for all of BioSimulator's conventions, including all of the features of the COMBINE archive format and SED-ML. To enable us to test tools involving a broad range of formats and algorithms, the test suite uses the specifications of tools to select appropriate archives for their validation from a corpus of curated archives and then uses these curated archives to computationally generate additional archives for testing specific aspects of BioSimulators' conventions. This design enables us to pinpoint issues with simulation tools, and it makes it easy to expand the test suite to additional model formats and methods. The test suite can be executed through a command-line interface or the GitHub issues deployment described below. More information about these validation tools is available in \textcolor{citecolor}{Section~S3}.

\subsection{Submission of simulation tools to the registry}
Developers can submit tools to the registry by submitting issues to the BioSimulators GitHub repository. Once an issue is created, GitHub actions is then used to execute the test suite described above, and any test failures are reported as messages to the issue. The first time a simulation tool passes the test suite, our team also manually reviews the capabilities of the tool and uses the issue to discuss any suggested revisions with the submitter. This manual review enables us to check aspects of tools that are challenging to test programmatically, such as the completeness of their specifications. This combination of machine and human review enables us to rigorously review each version of each tool with minimal effort.

We chose to use GitHub issues to manage the submission of simulation tools for two reasons. First, this enables the community to see how each tool was validated. Second, this provides developers an API for programmatically submitting tools. Importantly, this API makes it easy for developers to keep their tools up-to-date in BioSimulators. For example, developers can use this API within GitHub actions. Currently, half of the containerized tools registered with BioSimulators automatically release each version to BioSimulators. Third, GitHub issues enables our team to monitor problems that developers are encountering and help them. 

\section{Design, implementation, and deployment}
BioSimulators is composed of a set of conventions for consistently representing simulation projects and simulation tools; a set of tools for validating whether simulation projects and tools are consistent with these conventions; a collection of standardized Python APIs, command-line interfaces, and Docker images for simulation tools; a Docker image repository for these tools; a database for their specifications; a REST API for updating and querying this image repository and database; and a graphical user interface for browsing the database, validating projects, and getting recommendations for algorithms and tools.

The interfaces for simulators and the tools for validating simulation projects and simulators were primarily implemented with Python using libraries such as jlibSEDML, libCellML, libCOMBINE, libOmexMeta, libSBML, lib\-SED\-ML, pyBioNetGen, pyNeuroML, RBApy, Smoldyn, and XPP. The containerized interfaces for simulators were developed using Docker. The tools for validating logs of the execution of simulation projects and the specifications of simulation tools, the database of simulation tools, the REST API to the database, and the web application were implemented in TypeScript using NestJS, MongoDB, and Angular.

The database, API, web application, and test suite for simulation tools are deployed using Mongo Atlas, Google Cloud, Netlify, and GitHub, respectively. The containerized simulation tools are stored using GitHub Container Registry.

More information about the architecture, implementation, and deployment of BioSimulators is available in \textcolor{citecolor}{Section~S6}.

\section{Use cases}
\subsection{Sharing, reproducing, and reusing simulations}
We believe that BioSimulators makes it easier to share, reproduce, and reuse simulations by simplifying the installation and execution of simulators. Once an investigator has learned BioSimulators' conventions, they can run a broad range of simulations involving a variety of tools. In particular, we believe that simple web applications for using BioSimulators, such as runBioSimulations \cite{shaikh2021runbiosimulations}, will empower peer reviewers to review simulations more deeply, leading to better evaluation of modeling studies.

\subsection{Quality-controlling simulations}
We believe that BioSimulators' tool for integrated validation of simulation projects is excellent for identifying problems and other potential issues with simulations. For example, we are working with multiple model repositories to identify and correct issues in published simulation projects. More information is available in \textcolor{citecolor}{Section~S7}.

\subsection{Comparing simulation tools}
BioSimulators' registry of simulation tools is ideal for comparing and testing tools. In particular, by comparing the outputs of multiple tools, BioSimulators could help identify potential errors in tools. For example, BioSimulators has helped the BioNetGen, pyNeuroML, VCell, and other teams find and fix bugs in their tools.

\subsection{Multiscale simulation with multiple algorithms and tools}
By providing consistent Python APIs for simulation tools, we believe that BioSimulators makes it easier to combine multiple simulations of various subsystems and scales into multiscale simulations. In particular, BioSimulators makes it easier to combine simulations that require multiple model formats, simulation algorithms, and simulation tools. For example, the Vivarium Collective \cite{agmon2022vivarium} has begun to develop capabilities for co-simulating multiple BioSimulators tools.

\section{Discussion}
In summary, BioSimulators simplifies simulation by making it easier to find, obtain, and run appropriate tools for particular projects. Importantly, BioSimulators supports a broad range of simulation projects by using several formats and ontologies to encapsulate and abstract individual formats and tools, including model formats such as BNGL, SED-ML, KiSAO, the COMBINE archive format, HDF5, and Docker. We anticipate that BioSimulators will enhance several stages of the modeling life cycle. For example, we anticipate BioSimulators will encourage more reuse of published simulations by simplifying their execution, spur multiscale simulation by making it easier to combine multiple simulations of various subsystems, promote more predictive simulations by empowering peer reviewers to deeply review simulations, and stimulate higher quality simulation repositories by enabling more holistic validation of simulations. Below, we summarize how we plan to continue to enhance BioSimulators.

\subsection{Systemizing additional simulation domains}
Going forward, we aim to work with the community to expand the BioSimulators ecosystem to additional domains, including adding additional formats, frameworks, and algorithms to EDAM, SBO, and KiSAO;
developing conventions for using SED-ML with additional model formats; incorporating additional model formats into our simulation project validation suite; curating additional example COMBINE archives for our simulation tool test suite; and developing interfaces to additional simulation tools. Currently, we are working with the CoLoMoTo community to expand BioSimulators’ capabilities for logical modeling, such as calculations of state transition graphs and trap spaces.

\subsection{Accelerating more holistic simulation workflows}
By building on SED-ML, BioSimulators is currently limited to simple simulation workflows that consist of models, modification of models, the simulation of models, basic calculations of simulation results, exporting simulation results, and 2D line and 3D surface plots. In contrast, real-world studies often involve additional tasks, such as aggregating, normalizing, and integrating data from multiple sources; using this data to build and calibrate models; performing complex data reductions on simulation results; and generating a variety of visualizations of simulation results. Going forward, we aim to work with the community to develop a new version of SED-ML, which can capture a broader range of tasks, and develop a workflow engine that can use multiple containerized tools to modularly execute the individual tasks of these workflows. This design would also make it easier for software developers to participate in BioSimulators by lowering the responsibilities of tools from executing entire workflows to executing individual tasks.

\subsection{Enhanced recommendations of simulation methods}
Finally, we also aim to develop an additional wizard that helps novices identify appropriate formats, frameworks, algorithms, and tools for their work. Our current recommendation services require users to have advanced knowledge of simulation methodology. In contrast, we aim to develop a wizard that asks users questions about the systems and scales they would like to model and recommends appropriate formats, frameworks, algorithms, and tools. We anticipate that this would help more investigators model biology.



\section{Availability}
BioSimulators is freely available without registration at \href{https://biosimulators.org}{https://\allowbreak{}bio\allowbreak{}sim\allowbreak{}u\allowbreak{}la\allowbreak{}tors.\allowbreak{}org}. This website contains links to the simulation tools, REST API, examples, and documentation. The source code for BioSimulators is openly available under the MIT license. More information is available in \textcolor{citecolor}{Section~S9}.

\section{Supplementary data}
\textcolor{linkcolor}{Supplementary Data} are available online.


\section{Funding}
This work was supported by National Institutes of Health awards P41EB023912, R24GM137787, and R35GM119771.

\subsection{Conflict of interest statement.} None declared.


\begin{thebibliography}{10}

\bibitem{carrera2015build}
Carrera,J. and Covert,M.~W. (2015)
Why build whole-cell models?.
{\em Trends Cell Biol.,} {\bf 25}, 719--722.

\bibitem{marucci2020computer}
Marucci,L., Barberis,M., Karr,J., Ray,O., Race,P.~R., de~Souza~Andrade,M.,
  Grierson,C., Hoffmann,S.~A., Landon,S., Rech,E. et al. (2020)
Computer-aided whole-cell design: taking a holistic approach by integrating
  synthetic with systems biology.
{\em Front. Bioeng. Biotechnol.,} p. 942.

\bibitem{szigeti2018blueprint}
Szigeti,B. et al. (2018)
A blueprint for human whole-cell modeling.
{\em Curr. Opin. Syst. Biol.,} {\bf 7}, 8--15.

\bibitem{waltemath2016toward}
Waltemath,D., Karr,J.~R., Bergmann,F.~T., Chelliah,V., Hucka,M., Krantz,M.,
  Liebermeister,W., Mendes,P., Myers,C.~J., Pir,P. et al. (2016)
Toward community standards and software for whole-cell modeling.
{\em IEEE Trans. Biomed. Eng.,} {\bf 63}, 2007--2014.

\bibitem{ebrahim2013cobrapy}
Ebrahim,A., Lerman,J.~A., Palsson,B.~O. and Hyduke,D.~R. (2013)
{COBRApy}: constraints-based reconstruction and analysis for {Python}.
{\em BMC Syst. Biol.,} {\bf 7}, 1--6.

\bibitem{bergmann2017copasi}
Bergmann,F.~T., Hoops,S., Klahn,B., Kummer,U., Mendes,P., Pahle,J. and Sahle,S.
  (2017)
{COPASI} and its applications in biotechnology.
{\em J. Biotechnol.,} {\bf 261}, 215--220.

\bibitem{clerx2020cellml}
Clerx,M., Cooling,M.~T., Cooper,J., Garny,A., Moyle,K., Nickerson,D.~P.,
  Nielsen,P.~M. and Sorby,H. (2020)
{CellML} 2.0.
{\em J. Integr. Bioinform.,} {\bf 17}, 20200021.

\bibitem{keating2020sbml}
Keating,S.~M., Waltemath,D., K{\"o}nig,M., Zhang,F., Dr{\"a}ger,A.,
  Chaouiya,C., Bergmann,F.~T., Finney,A., Gillespie,C.~S., Helikar,T. et al.
  (2020)
{SBML Level} 3: an extensible format for the exchange and reuse of biological
  models.
{\em Mol. Syst. Biol.,} {\bf 16}, e9110.

\bibitem{malik2020biomodels}
Malik-Sheriff,R.~S., Glont,M., Nguyen,T.~V., Tiwari,K., Roberts,M.~G.,
  Xavier,A., Vu,M.~T., Men,J., Maire,M., Kananathan,S. et al. (2020)
{BioModels}—15 years of sharing computational models in life science.
{\em Nucleic Acids Res.,} {\bf 48}, D407--D415.

\bibitem{mcdougal2017twenty}
McDougal,R.~A., Morse,T.~M., Carnevale,T., Marenco,L., Wang,R., Migliore,M.,
  Miller,P.~L., Shepherd,G.~M. and Hines,M.~L. (2017)
Twenty years of {ModelDB} and beyond: building essential modeling tools for the
  future of neuroscience.
{\em J. Comput. Neurosci.,} {\bf 42}, 1--10.

\bibitem{Smith2021}
Smith,L.~P., Bergmann,F.~T., Garny,A., Helikar,T., Karr,J., Nickerson,D.,
  Sauro,H., Waltemath,D. and König,M. (2021)
{The Simulation Experiment Description Markup Language (SED-ML): language
  specification for Level 1 Version 4}.
{\em J. Integr. Bioinform.,} {\bf 18}, 20210021.

\bibitem{bergmann2014combine}
Bergmann,F.~T. et al. (2014)
{COMBINE archive and OMEX format: one file to share all information to
  reproduce a modeling project}.
{\em BMC Bioinformatics,} {\bf 15}, 1--9.

\bibitem{courtot2011controlled}
Courtot,M., Juty,N., Kn{\"u}pfer,C., Waltemath,D., Zhukova,A., Dr{\"a}ger,A.,
  Dumontier,M., Finney,A., Golebiewski,M., Hastings,J. et al. (2011)
Controlled vocabularies and semantics in systems biology.
{\em Mol. Syst. Biol.,} {\bf 7}, 543.

\bibitem{peters2017jws}
Peters,M., Eicher,J.~J., van Niekerk,D.~D., Waltemath,D. and Snoep,J.~L. (2017)
The {JWS Online} simulation database.
{\em Bioinformatics,} {\bf 33}, 1589--1590.

\bibitem{shaikh2021runbiosimulations}
Shaikh,B., Marupilla,G., Wilson,M., Blinov,M.~L., Moraru,I.~I. and Karr,J.~R.
  (2021)
{RunBioSimulations: an extensible web application that simulates a wide range
  of computational modeling frameworks, algorithms, and formats}.
{\em Nucleic Acids Res.,} {\bf 49}, W597--W602.

\bibitem{agmon2022vivarium}
Agmon,E., Spangler,R.~K., Skalnik,C.~J., Poole,W., Peirce,S.~M., Morrison,J.~H.
  and Covert,M.~W. (02, 2022)
Vivarium: an interface and engine for integrative multiscale modeling in
  computational biology.
{\em Bioinformatics,}
btac049.

\end{thebibliography}

\end{document}